\documentclass[%
 reprint,
superscriptaddress,
 amsmath,amssymb,
 aps,
 prb
]{revtex4-2}

\usepackage{physics}
\usepackage{braket}
\usepackage{qcircuit}
\usepackage{graphicx}
\usepackage{dcolumn}
\usepackage{bm}
\usepackage[hidelinks]{hyperref}
\usepackage{tensor}
\usepackage{xcolor}
\usepackage{threeparttable}
\newcommand{\brasub}[2]{\tensor[_{#2}]{\bra{#1}}{} }

\newcommand{\wtli}[1]{\textcolor{black}{#1}}

\begin{document}

\preprint{APS/123-QED}

\title{Efficient Quantum Simulation of Electron-Phonon Systems
by Variational Basis State Encoder 
}

\author{Weitang Li}
\affiliation{Tencent Quantum Lab, Tencent, Shenzhen, China}

\author{Jiajun  Ren}
\affiliation{College of Chemistry, Beijing Normal Univerisity, Beijing, China}

\author{Sainan Huai}
\affiliation{Tencent Quantum Lab, Tencent, Shenzhen, China}

\author{Tianqi Cai}
\affiliation{Tencent Quantum Lab, Tencent, Shenzhen, China}

\author{Zhigang Shuai}
\email{zgshuai@tsinghua.edu.cn}
\affiliation{Department of Chemistry, Tsinghua University, Beijing, China}
\affiliation{School of Science and Engineering, The Chinese University of Hong Kong, Shenzhen, China}

\author{Shengyu Zhang}
\email{shengyzhang@tencent.com}
\affiliation{Tencent Quantum Lab, Tencent, Shenzhen, China}

\date{\today}

\begin{abstract}
Digital quantum simulation of electron-phonon systems requires truncating infinite phonon levels into $N$ basis states
and then encoding them with qubit computational basis.
Unary encoding and binary encoding are the two most representative encoding schemes,
which demand $\mathcal{O}(N)$ and $\mathcal{O}(\log{N})$ qubits as well as $\mathcal{O}(N)$ and $\mathcal{O}(N\log{N})$ quantum gates respectively.
In this work, we propose a variational basis state encoding algorithm
that reduces the scaling of the number of qubits and quantum gates
to both $\mathcal{O}(1)$ \wtli{for systems obeying the area law of entanglement entropy}.
The cost for the scaling reduction is a constant amount of additional measurement.
The accuracy and efficiency of the approach are verified by both numerical simulation and realistic quantum hardware experiments.
In particular, we find using one or two qubits for each phonon mode is sufficient to produce quantitatively correct results 
across weak and strong coupling regimes.
Our approach paves the way for practical quantum simulation of electron-phonon systems on both near-term hardware and error-corrected quantum computers.

\end{abstract}

\maketitle

\section{Introduction}
Electron-phonon couplings are pervasive in quantum materials,
governing phenomena such as 
charge transport in semiconductors~\cite{Bardeen50},
vibrational spectra~\cite{Thewalt05}, 
polaron formation~\cite{Marcelo01},
and superconductivity~\cite{Pickett89}.
Classically, expensive numerical methods such as
density matrix renormalization group (DMRG) and quantum Monte-Carlo (QMC) are required
to accurately simulate electron-phonon systems due to the interior many-body interaction~\cite{White98,Devereaux12, Mishchenko15, Yao21, sous2021phonon, Li21, zhang2023bipolaronic}.
Quantum computers hold promise for the simulation of quantum systems
with exponential speedup over classical computers~\cite{lloyd1996universal}.
In the wake of the tremendous progress in the implementation of quantum computers~\cite{arute2019quantum, xu2022digital} 
and the dawning of the noisy intermediate-scale quantum (NISQ) era~\cite{Preskill18},
how to solve electron-phonon coupling problems with quantum computers
has attracted a lot of research interest~\cite{Harnik18, ollitrault2020nonadiabatic, jaderberg2022recompilation, lee2022variational, Shuai22, denner2023hybrid}.

A prominent problem for the digital quantum simulation of electron-phonon systems is how to encode
the infinite phonon states with finite quantum computational basis states.
The first step is usually truncating the infinite phonon states into $N$ basis states $\{\ket{m}\}$ 
and then the second step is encoding $\{\ket{m}\}$ into quantum computational basis $\{\ket{n}\}$.
The phonon basis states are usually the $N$ lowest harmonic oscillator eigenstates 
or $N$ uniformly distributed grid basis states.
There are two established strategies to perform the encoding $\{\ket{m}\} \mapsto \{\ket{n}\}$~\cite{sawaya2020resource, di2021improving}.
The first is unary encoding~\cite{somma2003quantum, miessen2021quantum},
in which each $\ket{m}$ is encoded to $\ket{00\dots 1_m \dots 00}$, and the total number of qubits required scales as $\order{N}$.
The second is binary encoding,
in which each $\ket{m}$ is encoded to $\prod_i \ket{\lfloor \frac{m}{2^i} \rfloor \mod 2}$ 
represented by $\order{\log{N}}$ qubits~\cite{Harnik18, ollitrault2020nonadiabatic, jaderberg2022recompilation}.
In terms of two-qubit gates required to simulate quantum operators such as $\hat b^\dagger \pm \hat b$ and $\hat b^\dagger \hat b$,
unary encoding scales as $\order{N}$ and binary encoding scales as $\order{N\log{N}}$~\cite{di2021improving}.
The features of unary encoding and binary encoding are summarized in Table 1.
Compared to the simulation of electrons, 
the simulation of phonons consumes quantum resources in a much faster manner,
which becomes the bottleneck for efficient quantum simulation of electron-phonon systems.

\begin{table}[h]
\caption{\label{tab:toc}%
Comparison of traditional encoding schemes and the proposed variational encoding
in terms of encoding formula, the number of qubits $N_{\rm{qubit}}$ required 
and the number of quantum gates $N_{\rm{gate}}$ required to simulate common phonon operators such as $\hat b^\dagger \pm \hat b$
and $\hat b^\dagger \hat b$ .
}
\begin{ruledtabular}
\begin{tabular}{cccc}
Scheme  &
Formula&
$N_{\rm{qubit}}$&
$N_{\rm{gate}}$\\
\colrule
Unary & $\ket{m} \mapsto \ket{00\dots 1_m \dots 00}$ & $\order{N}$ & $\order{N}$\\
Binary & $\ket{m} \mapsto \prod_i \ket{\lfloor \frac{m}{2^i} \rfloor \mod 2}$ & $\order{\log{N}}$ & $\order{N\log{N}}$\\
\bf{Variational} & $\sum_m C_{mn} \ket{m} \mapsto \ket{n}$ & $\order{1}$ & $\order{1}$ \\
\end{tabular}
\end{ruledtabular}
\end{table}

In this work, we propose a new basis encoding scheme called variational encoding.
Variational encoding maps linear combinations of $\ket{m}$ that are most entangled to the simulated system 
into the computational basis, i.e. $\sum_m C_{mn} \ket{m} \mapsto \ket{n}$,
where $C_{mn}$ is determined by variational principle.
The advantage of our approach is that, by encoding only the most entangled states and discarding the ones with little entanglement, 
the size of $\{\ket{n}\}$ can be made irrelevant to the size of $\{\ket{m}\}$.
In other words, the number of qubits required scales as $\order{1}$.
Consequently, the scaling for the number of gates is also $\order{1}$.
\wtli{
The premise of the scaling reduction is the area law of entanglement entropy.
}
Variational encoding is best suited to work in combination with variational quantum algorithms
such as variational quantum eigensolver (VQE)~\cite{peruzzo2014variational, mcclean2016theory} 
and variational quantum dynamics (VQD)~\cite{li2017efficient, yuan2019theory}.
Besides, the variational encoding is also compatible with 
Trotterized time evolution and quantum phase estimation (QPE)~\cite{lloyd1996universal, abrams1999quantum, aspuru2005simulated}.
Numerical simulation and experiments on realistic quantum hardware based on the Holstein model and spin-boson model
shows that using 1 or 2 qubits for each phonon mode is typically sufficient for highly accurate results even in the strong coupling regime.

\section{Variational Basis State Encoder}
In this section, we present a more rigorous formulation of the variational basis state encoder.
To encode each phonon mode $l$, encoded by $N_l$ qubits, we define the variational basis state encoder $\hat B[l]$ as follows
\begin{equation}
\label{eq:mapping}
    \hat B[l] := \sum_{m}\sum_{n=1}^{2^{N_l}} C[l]_{mn} \ket{n}_l \brasub{m}{l} \ ,
\end{equation}
with orthonormal constraint
\begin{equation}
    \hat B[l] \hat B[l]^\dagger = \hat I \ ,
\end{equation}
or equivalently
\begin{equation}
\label{eq:c-orthonormal}
    \sum_m C[l]_{mn}C[l]^*_{mn'}=\delta_{nn'} \ .
\end{equation}
\wtli{
In this paper, we use $\ket{m}$ to represent phonon states and $\ket{n}$ to represent qubit states.
Eq.~\ref{eq:mapping} can be rewritten as
\begin{equation}
    \hat B[l] = \sum_{n=1}^{2^{N_l}} \ket{n}_l \sum_{m} C[l]_{mn} \brasub{m}{l}
\end{equation}
and it is clear that $\hat B$ performs $\sum_m C_{mn} \ket{m} \mapsto \ket{n}$, 
}
The original Hamiltonian in $\ket{m}$ basis $\hat H$ can then be encoded to $\ket{n}$ basis using the following expression
\begin{equation}
\hat  {\tilde{H}} := \prod_l \hat B[l] \hat H \prod_l \hat B[l]^\dagger \ .   
\end{equation}

For both static and dynamic cases, encoder coefficients $C$ are determined by variational principle.
In the remainder of the section, we will derive the equation for $C$.
We use atomic units throughout the paper.

\subsection{Time-independent equation}
Suppose the quantum circuit is parameterized by $\ket{\phi} = \prod_k e^{i \theta_k \hat R_k} \ket{\phi_0}$,
and then the ground state Lagrangian with multipliers $\lambda_{lnn'}$ is
\begin{equation}
\label{eq:static-lag}
    \mathcal{L} = \braket{\phi|\hat{\tilde{H}}|\phi}  + \sum_{lnn'} \lambda_{lnn'} (\sum_m C[l]_{mn} C[l]_{mn'}^* - \delta_{nn'} ) \ .
\end{equation}
Taking the derivative with respect to $\theta_k$ immediately leads to traditional VQE with encoded Hamiltonian $\hat{\tilde{H}}$
\begin{equation}
\label{eq:static-vqe}
    \pdv{\braket{\phi|\hat{\tilde{H}}|\phi}}{\theta_k} = 0
\end{equation}
Taking the derivative with respect to $C[l]_{mn}$ and setting it to 0 yields
\begin{equation}
\label{eq:static-lag-derivative}
    \braket{\phi|n}_l \brasub{m}{l} \hat{\tilde{H}}'[l] \ket{\phi} + \sum_{n'} \lambda_{lnn'} C[l]_{mn'}^* = 0 \ ,
\end{equation}
where
$ \hat{\tilde{H}}'[l]$ is  the the half-encoded Hamiltonian
\begin{equation}
    \hat{\tilde{H}}'[l] := \prod_{k \neq l} \hat B[k] \hat H \prod_k \hat B[k]^\dagger \ .
\end{equation}
Multiply Eq.~\ref{eq:static-lag-derivative} with $C[l]_{mn''}$
and use the $C[l]$ orthonormal condition Eq.~\ref{eq:c-orthonormal} to get $\lambda$
\begin{equation}
\label{eq:static-lambda}
    \lambda_{lnn'} = -\sum_m C[l]_{mn'}\braket{\phi|n}_l \brasub{m}{l} \hat{\tilde{H}}'[l] \ket{\phi} \ .
\end{equation}
Define projector
\begin{equation}
    \hat P := \hat B[l]^\dagger B[l] = \sum_{mm'}\sum_{n} \ket{m}_l C[l]_{mn}^* C[l]_{m'n} \brasub{m'}{l} \ .
\end{equation}
Substitute $\lambda$ (Eq.~\ref{eq:static-lambda}) into Eq.~\ref{eq:static-lag-derivative} then yields
\begin{equation}
    \braket{\phi|n}_l \brasub{m}{l} \hat{\tilde{H}}'[l] \ket{\phi} 
    - \braket{\phi|n}_l \brasub{m}{l} \hat P \hat{\tilde{H}}'[l] \ket{\phi} = 0 \ .
\end{equation}
Rearranging and rewriting in matrix form, we get the equation for $C[l]$
\begin{equation}
\label{eq:static-c}
    (1 - \hat P[l]) \braket{\phi | \hat{\tilde{H}}'[l] | \phi} = 0 \ .
\end{equation}
Here $C[l]$ is contained in $\hat P[l]$ and $\hat{\tilde{H}}'[l] $.

To summarize, circuit parameters $\theta_k$ are solved by VQE according to Eq.~\ref{eq:static-vqe},
and variational parameters $C[l]$ are determined by solving Eq.~\ref{eq:static-c} classically.
Because Eq.~\ref{eq:static-vqe} contains $C[l]$ and Eq.~\ref{eq:static-c} contains $\theta_k$
, $\theta_k$ and $C[l]$ are solved iteratively until convergence.
In the following, this iteration is termed macro-iteration to avoid confusion with VQE iteration.

\subsection{Quantum circuit measurement}
In this section, we discuss the quantum circuit measurement required to solve $C[l]$ from Eq.~\ref{eq:static-c}.
The key quantity to be computed is matrix $G[l]_{mn}$, defined as
\begin{equation}
    G[l]_{mn} := \braket{\phi|n}_l \brasub{m}{l} \hat{\tilde{H}}'[l] \ket{\phi} \ .
\end{equation}
Suppose the Hamiltonian can be written as a sum of direct products
\begin{equation}
\begin{aligned}
    \hat H & = \sum_x^{M} \hat h_x \ , \\
    \hat h_x & = \prod_k \hat h[k]_x
\end{aligned}
\end{equation}
where $M$ is the total number of terms in the Hamiltonian and $\hat h[k]_x$ acts on the $k$th degree of freedom.
Similar to the encoded Hamiltonian, the encoded local operator is denoted as $\hat{\tilde{h}}[k]_x$
\begin{equation}
    \hat{\tilde{h}}[k]_x := \hat B[k] \hat{h}[k]_x \hat B[k]^\dagger \ .
\end{equation}
For electron degree of freedom a dummy encoder $\hat B[k] = \hat I$ is used for notational simplicity.
$G[l]_{mn}$ can then be written as
\begin{equation}
        G[l]_{mn} = \sum_x^M \braket{\phi|n}_l \brasub{m}{l} 
    \hat{h}[l]_x \hat B[l]^\dagger \prod_{k \neq l} \hat{\tilde{h}}[k]_x \ket{\phi} \ .
\end{equation}
Next, represent $\hat h[l]_x$ in operator form
\begin{equation}
    \hat h[l]_x = \sum_{mm'} h[l]_{xm'm} \ket{m'}_l \brasub{m}{l} \ .
\end{equation}
$G[l]_{mn}$ then becomes
\begin{equation}
\begin{aligned}
     G[l]_{mn} & = \sum_x^M \sum_{m'n'} h[l]_{xmm'} C[l]_{m'n'} J[l]_{xnn'} \ , \\
     J[l]_{xnn'} & := \braket{\phi|n}_l 
     \brasub{n'}{l}\prod_{k \neq l} \hat{\tilde{h}}[k]_x \ket{\phi} \ .
\end{aligned}
\end{equation}
Thus to evaluate $G[l]_{mn}$ it is sufficient to measure $J[l]_{xnn'}$.
$\ket{n}_l \brasub{n'}{l}$ in general is not Hermitian,
but the real and imaginary parts can be measured separately with $ (\ket{n}_l \brasub{n'}{l} + \ket{n'}_l \brasub{n}{l})$
and $i(\ket{n'}_l \brasub{n}{l} - \ket{n}_l \brasub{n'}{l})$.

Assuming the number of measurement shots for each Pauli string is $N_{\rm{shots}}$,
the number of measurements to determine $J[l]$ is thus $\order{2^{N_l}M N_{\rm{shots}}}$,
which is polynomial to the system size and does not increase with $N$.
\wtli{
After $J[l]$ is measured,
evaluating $G[J]$ and the left-hand side of Eq.~\ref{eq:static-c} scales as $\order{2^{N_l}N^2 M}$ by matrix multiplication on a classical computer. 
Considering the measurement of a parameterized quantum circuit takes much longer time than a float-point number operation on classical computers,
the classical workload is negligible compared to the additional measurements for reasonable values of $N_{\rm{shots}}$ and $N$, such as  $N_{\rm{shots}}=4096$ and $N=64$.
Thus the reduction in quantum resources is not achieved by increasing classical resources~\cite{ryabinkin2020iterative}.
}
If the number of phonon modes is assumed to be linear with $M$
and each $C[l]$ is updated independently, then the total number of measurements
for all $C[l]$ is $\order{2^{N_l}M^2 N_{\rm{shots}}}$.
The measurement overhead increases exponentially with $N_l$.
Due to arguments presented later, $N_l$ is usually small and does not increase with system size.
From numerical experiments, we find $N_l \le 2$ is sufficient to produce excellent results.

\subsection{Time-dependent equation}
For time-dependent problems, it is convenient to define 
\begin{equation}
    \ket{ \psi} := \prod_l \hat B[l]^\dagger \ket{\phi}
\end{equation}
and use $\Theta_K$ denote both $\theta_k$ and $C[l]$.
The Lagrangian with multipliers $\lambda_{lnn'}$ and $\gamma_{lnn'}$ is then
\begin{equation}
\label{eq:dynamic-lag}
\begin{aligned}
    \mathcal{L} & = |i \sum_K \pdv{\ket{\psi}}{\Theta_K} \dot \Theta_K - \hat H \ket{\psi}|^2 \\
    & \quad + \sum_{lnn'} \lambda_{lnn'} \Re{ \sum_m C[l]^*_{mn} \dot C[l]_{mn'} } \\
    & \quad + \sum_{lnn'} \gamma_{lnn'} \Im{ \sum_m C[l]^*_{mn} \dot C[l]_{mn'} } \ .
\end{aligned}
\end{equation}
The constraints ensure that $C[l]_{mn}$ remains orthonormal during time evolution.
Taking the derivative with respect to $\dot \Theta_K$ yields
\begin{equation}
\begin{aligned}
\label{eq:dynamics-lag-deriv}
    \pdv{\mathcal{L}}{\dot \Theta_K} & = \sum_J \pdv{\bra{\psi}}{\Theta_J}\pdv{\ket{\psi}}{\Theta_K} \dot \Theta_J
    + \sum_J \pdv{\bra{\psi}}{\Theta_K}\pdv{\ket{\psi}}{\Theta_J} \dot \Theta_J \\
    & \quad + i\pdv{\bra{\psi}}{\Theta_K} \hat H \ket{\psi} - i\bra{\psi}\hat H \pdv{\ket{\psi}}{\Theta_K} \\
    & \quad + \sum_{lnn'} \lambda_{lnn'} \Re{ \sum_m C[l]^*_{mn} \pdv{\dot C[l]_{mn'}}{\dot \Theta_K} } \\
    & \quad + \sum_{lnn'} \gamma_{lnn'} \Im{ \sum_m C[l]^*_{mn} \pdv{\dot C[l]_{mn'}}{\dot \Theta_K} } \ .
\end{aligned}
\end{equation}
The subsequent derivation involves a more intricate process similar to  that of Eq.~\ref{eq:static-c}.
Elaborate details are documented  in Appendix~\ref{sec:app-theory}. Here, we provide an outline of the crucial steps.
First, consider the case where $\Theta_K = \theta_k$
and we find that
the equation of motion for $\theta_k$ is the same as vanilla VQD with encoded Hamiltonian $\hat{\tilde{H}}$
\begin{equation}
\label{eq:eom-theta}
\sum_j \Re{ \pdv{\bra{\phi}}{\theta_k} \pdv{\ket{\phi}}{\theta_j}  } \dot \theta_j = \Im{  \pdv{\bra{\phi}}{\theta_k}  \hat{\tilde{H}} \ket{\phi} }       \ .
\end{equation}
Next, consider the case where $\Theta_K = C[l]$,
which ultimately leads to the following equation for $\dot C[l]$
\begin{equation}
\label{eq:eom-c}
    i \rho[l] \dot C[l]^*
    = (1 - \hat P[l]) \braket{\phi | \hat{\tilde{H}}'[l] | \phi} \ ,
\end{equation}
where $\rho[l]_{nn'} = \Tr{\braket{\phi|n}_l \brasub{n'}{l}\phi \rangle}$ is the reduced density matrix for the $N_l$ qubits of $\ket{\phi}$.
Eq.~\ref{eq:static-c} represents a $\dot C[l] = 0$ stationary point during real and imaginary time evolution. The measurement cost is the same as the ground state algorithm.

While we have relied on parameterized quantum circuits (PQC) for our derivation thus far, it is worth noting that incorporating the variational encoder into Trotterized time evolution and QPE is a straightforward extension.
The VQD step described by Eq.~\ref{eq:eom-theta} can be naturally replaced by a Suzuki-Trotter time evolution step
$e^{-i\hat{\tilde{H}} \tau} \approx \prod_x^M e^{-i \hat{\tilde{h}}_x \tau}$ on a digital quantum simulator,
so that Hamiltonian simulation is performed via Trotterized time evolution instead of VQD.
To update $C[l]$ based on Eq.~\ref{eq:eom-c}, measurements on the circuit
should be performed for every or every several Trotter steps.
The variationally encoded ground state can then be  prepared by adiabatic state preparation,
whose energy is accessible by QPE using $\hat{\tilde{H}}$.

\subsection{Variational basis state encoder as an ansatz}
It is instructive to observe that if the variational basis encoder
is viewed as a wavefunction ansatz $\ket{\psi}$,
then the algorithm proposed in this work can be viewed as a generalization 
for the local basis optimization method for DMRG~\cite{zhang1998density, guo2012critical},
or a special case of the recently proposed quantum-classical hybrid tensor network~\cite{yuan2021quantum}.
Thus, $\hat B[l]$ captures the $2^{N_l}$ phonon states that are most entangled with the rest of the system.
For local Hamiltonian obeying the area law, the entanglement entropy between one phonon mode 
and the rest of the system $S$ is a constant~\cite{eisert2010colloquium}.
Consequently, $|\braket{\psi|\Psi}|^2$, 
the fidelity between the approximated encoded state and the target state  
has a lower bound of $\frac{2^{N_l}}{e^S}$,
which lays the theoretical foundation for the effectiveness of the variational encoding approach 
to ground state and low-lying excited state problems.

\section{Simulations}
\subsection{Numerical simulation on a noiseless simulator} 
The variational basis state encoder is first tested 
for VQE simulation of the one-dimensional Holstein model~\cite{holstein1959studies1, holstein1959studies2}
\begin{equation}
\label{eq:ham-holstein}
    \hat H = -\sum_{\langle i, j\rangle}V \hat a^\dagger_i \hat a_j + \sum_{i} \omega \hat b^\dagger_i \hat b_i 
    + \sum_{i} g \omega \hat a^\dagger_i \hat a_i (\hat b^\dagger_i + \hat b_i) \ ,
\end{equation}
where \wtli{$\hat a$ and $\hat b$ are annihilation operators for electron and phonon respectively},
$V$ is the hopping coefficient, $\langle i, j\rangle$ denotes nearest neighbour pairs with periodic boundary condition,
$\omega$ is the vibration frequency and $g$ is dimensionless coupling constant.
In the following, we assume $V=\omega=1$ and adjust $g$ for different coupling strengths.
We consider a three-site system corresponding to $3(N_l+1)$ qubits.
We use binary encoding to represent traditional encoding approaches. 
Unary encoding is expected to produce similar results with binary encoding only with different quantum resource budgets.
We devise the following ansatz
\begin{equation}
\label{eq:ansatz}
    \ket{\phi} = \prod_l^L \left \{ \prod_{\langle j, k\rangle} e^{\theta_{ljk} (\hat a^\dagger_j \hat a_k - \hat a^\dagger_k \hat a_j)}
    \prod_j e^{\theta_{lj}\hat a^\dagger_j \hat a_j (\hat b^\dagger_j - \hat b_j)} \right \}\ket{\phi_0} \ .
\end{equation}
where $L$ is the number of layers and $L=3$ is adopted.
More details of the simulation are included in the Appendix~\ref{sec:app-numerical}. 

\begin{figure}[t]
\includegraphics[width=\linewidth]{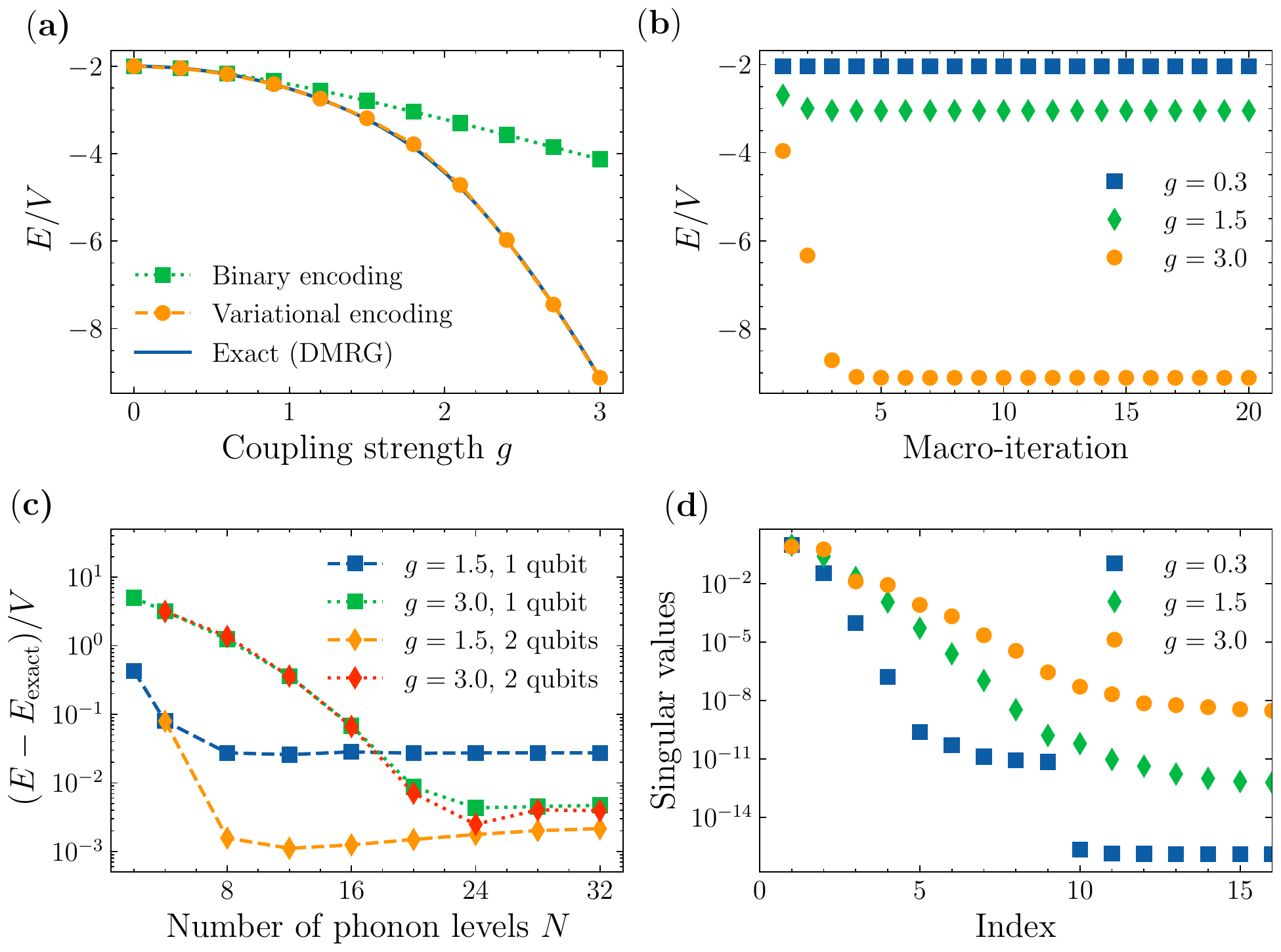}
\caption{\label{fig:holstein} 
Numerical simulation results for the ground state of the Holstein model.
(a) Ground state energy by  binary encoding and variational encoding with different coupling strength $g$;
(b) Convergence of ground state energy with respect to the macro-iteration for variational encoding;
(c) Ground state energy error for the variational encoding method at different numbers of phonon basis states $N$;
(d) The singular values for the Schmidt decomposition between the last phonon mode and the rest of the system.
}
\end{figure}

We first compare the accuracy of the variational encoding and the binary encoding with $N_l=1$. 
It is clear from Fig.~\ref{fig:holstein}(a) that
variational encoding is significantly more accurate than binary encoding, especially at the strong coupling regime.
Within the setup, binary encoding uses only two phonon basis states to describe each phonon mode, yet
the variational encoding is allowed to use up to 32 phonon basis states before combining them into the most entangled states.
We note that the quantum circuit used for variational encoding and binary encoding is essentially the same.
The number of macro-iterations to determine $C[l]$ is found to be rather small, as shown in Fig.~\ref{fig:holstein}(b).
Fully converged results are obtained within 5 iterations.
In Fig.~\ref{fig:holstein}(c) we show more details of the error for the variational approach.
The simulation error typically decreases exponentially with respect to the number of phonon levels $N$ included in $C[l]$.
It is worth noting that quantum computational resources, including the number of qubits, the number of gates in the circuit, and the number of measurements remain constant when $N$ is increased from 2 to 32.
Furthermore, by using 2 qubits to encode each mode, it is possible to further reduce the error 
at the $N \rightarrow \infty$ limit.
When $g = 3.0$, the error is not sensitive to $N_l$, which implies that the error is dominated by other sources
such as limitations of the ansatz, instead of the small $N_l$. 
Fig.~\ref{fig:holstein}(d) shows the singular values for the Schmidt decomposition between the last phonon mode and the rest of the system by DMRG. The exponential decay ensures the fast convergence of $N_l$.
The von Neumann entropy $S$ for the three systems is found to be 0.01, 0.25, and 0.65 respectively.
We also note the $g=1.5$ case has the largest 3rd singular value, which explains why setting $N_l=2$ significantly reduces the $g=1.5$ error in Fig.~\ref{fig:holstein}(c).

We now turn to 
the spin-relaxation dynamics of the spin-boson model~\cite{leggett1987dynamics}, described by the Hamiltonian
\begin{equation}
    \hat H = \frac{\epsilon}{2} \hat \sigma_z + \Delta \hat \sigma_x + \sum_j g_j \omega_j \hat \sigma_z (\hat b^\dagger_j + \hat b_j) 
    + \sum_j \omega_j \hat b^\dagger_j \hat b_j \ .
\end{equation}
where $\epsilon$ is the eigenfrequency and $\Delta$ is the tunneling rate.
The coupling term has a similar form with Eq.~\ref{eq:ham-holstein}
and is more commonly written as  $\sum_j c_j \hat \sigma_z \hat x_j$.
For systems in the condensed phase the coupling is usually characterized by the spectral density function
$\mathcal{J}(\omega) = \frac{\pi}{2}\sum_j \frac{c_j^2}{\omega_j} \delta(\omega - \omega_k)$.
In the following we assume $\epsilon = 0$ and $\Delta = 1$.
We first use VQD for the simulation and discuss Trotterized time evolution at last.
The variational Hamiltonian ansatz~\cite{wecker2015progress} with three layers is used if not otherwise specified.

\begin{figure}[t]
\includegraphics[width=\linewidth]{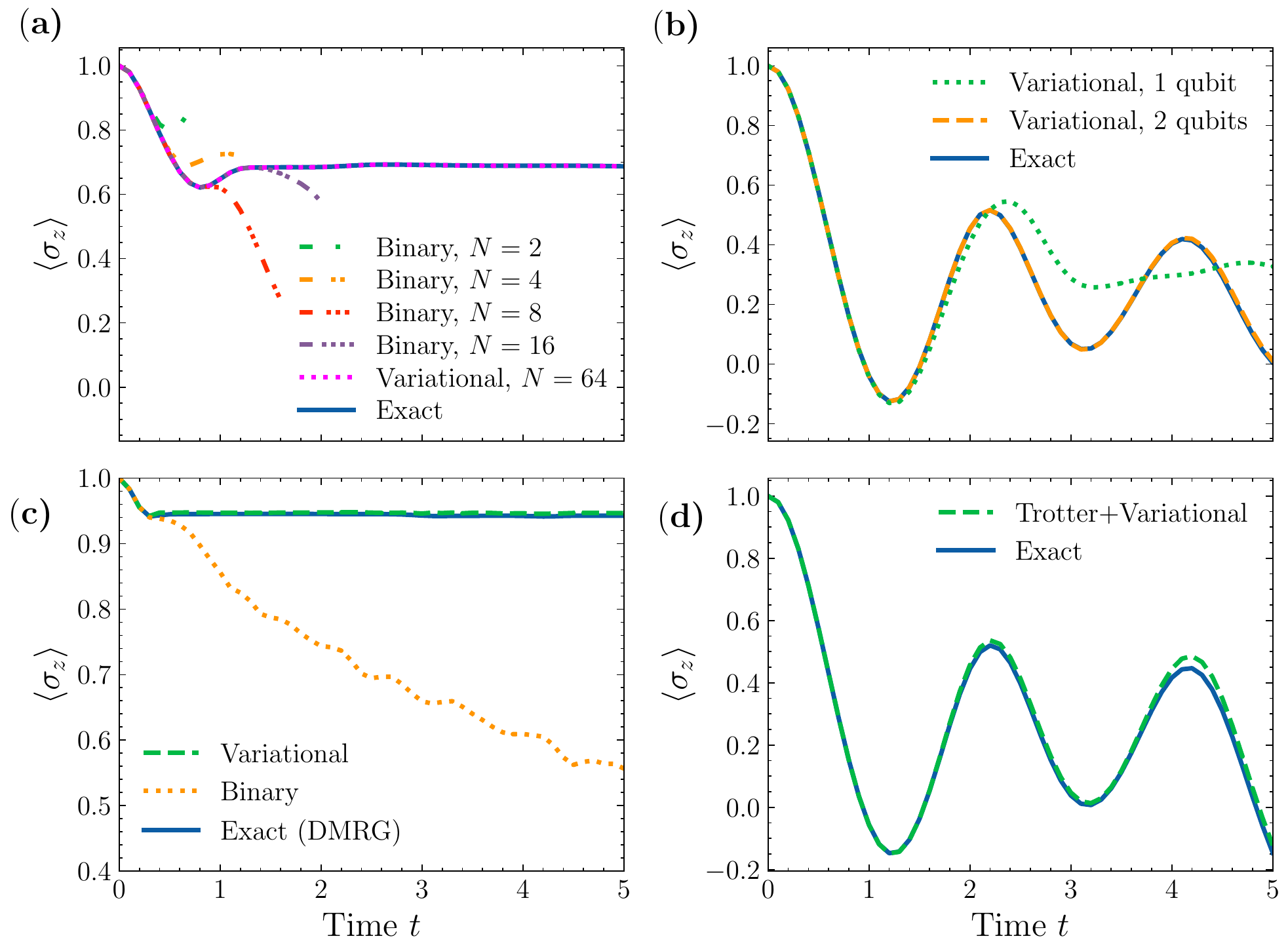}
\caption{\label{fig:sbm} 
Numerical simulation results for the spin-relaxation dynamics of the spin-boson model.
(a) Comparison between binary encoding with different numbers of phonon basis states and variational encoding for a one-mode spin-boson model;
(b) Variational encoding with different numbers of encoding qubits $N_l$ for a two-mode spin-boson model;
(c) Comparison between binary encoding and variational encoding for an 8-mode spin-boson model with sub-Ohmic spectral density;
(d) Trotterized time evolution with variational encoding based on a one-mode spin-boson model.
}
\end{figure}

The performance of variational encoding and binary encoding is first compared
based on a one-mode spin-boson model at the strong coupling  ($\omega=1$ and $g=3$) regime,
shown in Fig.~\ref{fig:sbm}(a).
Variational encoding with $N_l=1$ generates 
much more accurate dynamics than binary encoding with fewer qubits and quantum gates.
The simulation of binary encoding with $N_l > 4$ is prohibited by the deep circuit depth in the ansatz.
The variational encoding scheme is exceptionally efficient for this one-mode model
because Schmidt decomposition guarantees that 2 variational bases for the phonon mode 
are sufficient to exactly represent the system.
In Fig.~\ref{fig:sbm}(b) a two-mode model with $\omega_j=\frac{1}{2}, 1$ and $g_j=\frac{1}{2}, 1$ is used.
Variational encoding with $N_l=1$ is accurate at $t < 2$ but as the entanglement builds up 
the dynamics deviate from the exact solution.
Increasing $N_l$ to 2 effectively eliminates the error.
Next, we move on to a more challenging model with 8 modes,
in which
$\omega$ and $g$ are determined by
discretizing a sub-Ohmic spectral density $\mathcal{J}(\omega)=\frac{\pi}{2}\alpha \omega^s \omega_c^{1-s}e^{-\omega/\omega_c}$
following the prescription in the literature~\cite{wang2010coherent}.
The parameters are $s=\frac{1}{4}$, $\omega_c=4$ and $\alpha=10$.
As illustrated in Fig.~\ref{fig:sbm}(c) variational encoding with $N_l=1$ captures the localization behavior
yet binary encoding with $N_l=1$ completely fails.
The number of layers in the variational Hamiltonian ansatz is 8 and 32 for variational and binary encoding respectively.
Fig.~\ref{fig:sbm}(d) demonstrates the possibility to incorporate variational basis state encoder into Trotterized time evolution
with $\omega=g=1$ and $N_l=1$.
The measurement and the evolution of $C[l]$ are performed at each Trotter step.

\begin{figure}[t]
\includegraphics[width=\linewidth]{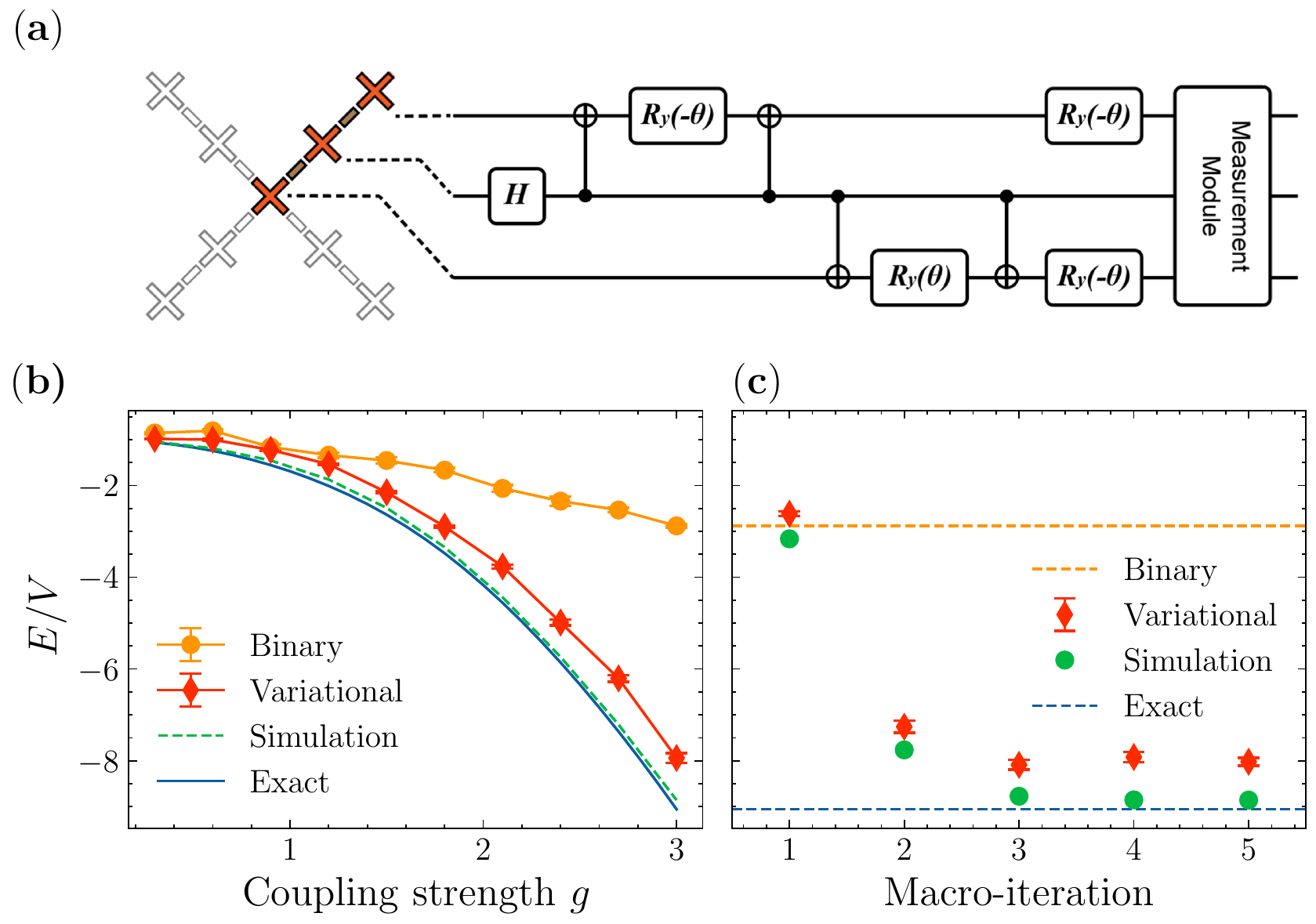}
\caption{\label{fig:hw} 
Quantum hardware experiments for the ground state energy of the Holstein model with variational basis state encoder.
(a) 3 qubits out of 9 qubits of a superconducting quantum processor and a one-parameter circuit are used for the simulation;
(b) Ground state energy by binary encoding and variational encoding;
(c) Convergence of ground state energy with respect to the macro-iteration for variational encoding. 
}
\end{figure}

\subsection{Verification on a superconducting quantum processor}
In this section, we verify the accuracy and efficiency of the variational encoder approach on a superconducting quantum processor~\cite{yan2018tunable, li2020tunable}.
We consider the ground state problem of a 2-site Holstein model described by Eq.~\ref{eq:ham-holstein} with $g=3$ and $N_l=1$.
The two electronic sites are represented by 1 qubit and the total number of qubits for the system is thus 3.
The quantum circuit for the simulation is depicted in Fig.~\ref{fig:hw}(a).
The electronic degree of freedom is mapped to the second qubit, and the two phonon modes are mapped to the first and the third qubits respectively.
There is one parameter to be determined by VQE in the circuit and 
the same ansatz is used for both binary encoding and variational encoding.
More simulation details can be found in the Supplemental Material.
In Fig.~\ref{fig:hw}(b) we show the ground state energy by variational encoding from weak to strong coupling, in analog to Fig.~\ref{fig:holstein}(a).  
The simulator result is based on the parameterized quantum circuit described in Fig.~\ref{fig:hw}(a)
without considering gate noise and measurement uncertainty.
The results in Fig.~\ref{fig:hw}(b) are consistent with that in Fig.~\ref{fig:holstein}(a).
The residual error is dominated by the intrinsic gate noise in the quantum computer. 
In Fig.~\ref{fig:hw}(c) we show the convergence with respect to the macro-iteration for variational encoding.
The algorithm is resilient to the presence of quantum noise and measurement uncertainty.
The convergent energy is reached within 5 iterations.

\section{Conclusion}
We proposed a variational basis state encoder to encode phonon basis states into quantum computational states
for efficient quantum simulation of electron-phonon systems.
The proposed variational encoding approach requires only $\order{1}$ qubits and $\order{1}$ quantum gates
\wtli{for systems obeying the area law of entanglement entropy},
which is significantly better than traditional encoding schemes
and enables quantum simulation of electron-phonon systems with smaller quantum processors and shallower circuits.
The additional measurement required to implement the approach is found to be also 
$\order{1}$ with respect to the number of phonon basis states
and it scales quadratically with the number of Pauli strings in the Hamiltonian.
The accuracy of the approach is ensured by the finite entanglement entropy between one phonon mode 
and the rest of the system in common electron-phonon systems.
The variational basis state encoder most naturally works with variational quantum algorithms
and is compatible with Trotterized time evolution, adiabatic state preparation, and QPE.
Numerical simulation and quantum hardware experiments 
based on VQE of the Holstein model and dynamics of the spin-boson model
indicate that variational encoding is more accurate and resource-efficient than traditional encoding methods.
In particular, using one or two qubits to represent each phonon mode is sufficient for accurate simulation
even at the strong coupling regime where $N=32$ phonon basis states are involved. 
The approach could also be extended to other quantum simulation problems involving an infinite or large local Hilbert space.

\section*{Acknowledgements}
We thank Jinzhao Sun and Shixin Zhang for helpful discussions.
This work is supported by the National Natural Science Foundation of China through grant numbers 22273005 and 21788102.
This work is also supported by Shenzhen Science and Technology Program.

\appendix
\begin{widetext}
\section{Derivation of time-dependent equation}
\label{sec:app-theory}
In this section, we derive the time-dependent equation for $C[l]$.
For time-dependent problems, $C[l]$ in general is complex
\begin{equation}
    C[l] = D[l] - iE[l] \ ,
\end{equation}
where both $D[l]$ and $E[l]$ are real.
The minus sign is for convenience expressing $\hat B^\dagger \ket{\phi}$.
From the definition we have
\begin{equation}
\label{eq:de-relation}
    \pdv{\ket{\psi}}{E[l]_{mn}} = i\pdv{\ket{\psi}}{D[l]_{mn}} \ .
\end{equation}

The starting point is Eq.~\ref{eq:dynamics-lag-deriv}
We first consider the case of $\Theta_K = \theta_k$, and then
\begin{equation}
\begin{aligned}
    \pdv{\mathcal{L}}{\dot \theta_k} & = \sum_J \pdv{\bra{\psi}}{\Theta_J}\pdv{\ket{\psi}}{\theta_k} \dot \Theta_J
    + \sum_J \pdv{\bra{\psi}}{\theta_k}\pdv{\ket{\psi}}{\Theta_J} \dot \Theta_J \\
    & \quad + i\pdv{\bra{\psi}}{\theta_k} \hat H \ket{\psi} - i\bra{\psi}\hat H \pdv{\ket{\psi}}{\theta_k} \\
    & = 2 \sum_J \Re{\pdv{\bra{\psi}}{\theta_k}\pdv{\ket{\psi}}{\Theta_J}} \dot \Theta_J - 
    2 \Im{\pdv{\bra{\psi}}{\theta_k} \hat H \ket{\psi}} \ ,
\end{aligned}
\end{equation}
which means at the $ \pdv{\mathcal{L}}{\dot \theta_k}=0$ minimum, we have
\begin{equation}
    \sum_J \Re{\pdv{\bra{\psi}}{\theta_k}\pdv{\ket{\psi}}{\Theta_J}} \dot \Theta_J = \Im{\pdv{\bra{\psi}}{\theta_k} \hat H \ket{\psi}}    \ .
\end{equation}
Substitute $\Theta_J$ with $\theta_k$, $D[l]_{mn}$ and $E[l]_{mn}$
\begin{equation}
\begin{aligned}
     \sum_J \Re{\pdv{\bra{\psi}}{\theta_k}\pdv{\ket{\psi}}{\Theta_J}} \dot \Theta_J & = 
     \sum_j \Re{\pdv{\bra{\psi}}{\theta_k}\pdv{\ket{\psi}}{\theta_j}} \dot \theta_j \\
     & \quad + \sum_{lmn} \Re{\pdv{\bra{\psi}}{\theta_k}\pdv{\ket{\psi}}{D[l]_{mn}}} \dot D[l]_{mn} \\
     & \quad + \sum_{lmn} \Re{\pdv{\bra{\psi}}{\theta_k}\pdv{\ket{\psi}}{E[l]_{mn}}} \dot E[l]_{mn} \ .
\end{aligned}
\end{equation}
Using Eq.~\ref{eq:de-relation} the last two terms become
\begin{equation}
    \sum_{lmn} \Re{\pdv{\bra{\psi}}{\theta_k}\pdv{\ket{\psi}}{D[l]_{mn}}} \dot D[l]_{mn}
    + \sum_{lmn} \Re{\pdv{\bra{\psi}}{\theta_k}\pdv{\ket{\psi}}{E[l]_{mn}}} \dot E[l]_{mn}
    = \sum_{lmn} \Re{\pdv{\bra{\psi}}{\theta_k}\pdv{\ket{\psi}}{D[l]_{mn}} \dot C[l]_{mn}^* } \ ,
\end{equation}
which is zero because
\begin{equation}
    \sum_{mn} \pdv{\bra{\psi}}{\theta_k}\pdv{\ket{\psi}}{D[l]_{mn}} \dot C[l]_{mn}^* 
    = \sum_{mn} \pdv{\bra{\phi}}{\theta_k}  \hat B[l] \ket{m}_l \brasub{n}{l} \dot C[l]_{mn}^*  \ket{\phi} = 0 \ ,
\end{equation}
where the constraint $\sum_m C[l]_{mn} \dot C[l]_{mn'}^* = 0$ is used. 
Thus the simplified equation of motion reads
\begin{equation}
    \sum_j \Re{\pdv{\bra{\psi}}{\theta_k}\pdv{\ket{\psi}}{\theta_j}} \dot \theta_j
    = \Im{\pdv{\bra{\psi}}{\theta_k} \hat H \ket{\psi}} \ ,
\end{equation}
or equivalently
\begin{equation}
\sum_j \Re{ \pdv{\bra{\phi}}{\theta_k} \pdv{\ket{\phi}}{\theta_j}  } \dot \theta_j 
= \Im{  \pdv{\bra{\phi}}{\theta_k} \hat{\tilde{H}} \ket{\phi} }       \ .
\end{equation}
In short, 
the equation of motion for $\theta_k$ is the same as vanilla VQD with encoded Hamiltonian $\hat{\tilde{H}}$ .

Next we consider the case of $\Theta_K = D[l]$ and $\Theta_K = E[l]$. After some complex algebra, we have
\begin{equation}
\label{eq:deriv-d}
    i \sum_{J} \pdv{\bra{\psi}}{D[l]_{mn}}\pdv{\ket{\psi}}{\Theta_J} \dot \Theta_J 
    + i \frac{1}{2} \sum_{n'} \lambda_{ln'n} C[l]_{mn'}^*
     - \frac{1}{2} \sum_{n'} \gamma_{ln'n}  C[l]_{mn'}^*
    = \pdv{\bra{\psi}}{D[l]_{mn}} \hat H \ket{\psi} \ .
\end{equation}
Similar to the case of $\Theta_K = \theta_k$, substitute $\Theta_J$  with $\theta_k$, $D[l]_{mn}$ and $E[l]_{mn}$
\begin{equation}
\begin{aligned}
    \sum_J \pdv{\bra{\psi}}{D[l]_{mn}}\pdv{\ket{\psi}}{\Theta_J} \dot \Theta_J & = 
    \sum_k \pdv{\bra{\psi}}{D[l]_{mn}}\pdv{\ket{\psi}}{\theta_k} \dot \theta_k 
     + \sum_{km'n'} \pdv{\bra{\psi}}{D[l]_{mn}}\pdv{\ket{\psi}}{D[k]_{m'n'}} \dot C[k]_{m'n'}^* \\
    & = \sum_k \pdv{\bra{\psi}}{D[l]_{mn}}\pdv{\ket{\psi}}{\theta_k} \dot \theta_k 
     + \sum_{n'} \pdv{\bra{\psi}}{D[l]_{mn}}\pdv{\ket{\psi}}{D[l]_{mn'}} \dot C[l]_{mn'}^* \ .
\end{aligned}
\end{equation}
Here the orthonormal condition is again used.
Substitute the equation back into Eq.~\ref{eq:deriv-d}.
\begin{equation}
\label{eq:tmp}
     i\sum_k \pdv{\bra{\psi}}{D[l]_{mn}}\pdv{\ket{\psi}}{\theta_k} \dot \theta_k
     + i\sum_{n'} \pdv{\bra{\psi}}{D[l]_{mn}}\pdv{\ket{\psi}}{D[l]_{mn'}} \dot C[l]_{mn'}^*
     + \frac{1}{2}\sum_{n'}(i\lambda_{ln'n} - \gamma_{ln'n})C[l]_{mn'}^* = \pdv{\bra{\psi}}{D[l]_{mn}} \hat H \ket{\psi} \ ,
\end{equation}

Following the same strategy with the derivation of  the time-independent equation, multiply Eq.~\ref{eq:tmp} with $C[l]_{mn}$
\begin{equation}
     i\sum_k \braket{\phi | n}_l \brasub{n'}{l} \pdv{\ket{\phi}}{\theta_k} \dot \theta_k
     + \frac{1}{2}(i\lambda_{ln'n} - \gamma_{ln'n})= \sum_m C[l]_{mn'}  \pdv{\bra{\psi}}{D[l]_{mn}} \hat H \ket{\psi} \ , 
\end{equation}
where $\sum_{m} C[l]^*_{mn'} C[l]_{mn} = \delta_{n'n}$ and  $\sum_{m} \dot C[l]^*_{mn'} C[l]_{mn} = 0$
are used.
Then, multiply again with $C[l]_{mn}^*$
\begin{equation}
    i\sum_k \pdv{\bra{\psi}}{D[l]_{mn}}\pdv{\ket{\psi}}{\theta_k} \dot \theta_k
     + \frac{1}{2}\sum_{n'}(i\lambda_{ln'n} - \gamma_{ln'n})C[l]_{mn'}^* = \hat P[l] \pdv{\bra{\psi}}{D[l]_{mn}} \hat H \ket{\psi} \ .
\end{equation}
Use this equation to eliminate $\lambda$ and $\gamma$ in Eq.~\ref{eq:tmp}, we get the equation of motion for $C[l]$
\begin{equation}
    i\sum_{n'} \pdv{\bra{\psi}}{D[l]_{mn}}\pdv{\ket{\psi}}{D[l]_{mn'}} \dot C[l]_{mn'}^*
    = (1 - \hat P[l])\pdv{\bra{\psi}}{D[l]_{mn}} \hat H \ket{\psi} \ ,
\end{equation}
which can be simplified to
\begin{equation}
    i \rho[l] \dot C[l]^*
    = (1 - \hat P[l]) \braket{\phi | \hat{\tilde{H}}'[l] | \phi} \ ,
\end{equation}
The measurement required for time evolution is in the same order as the static VQE algorithm.

In the end, we note that imaginary time evolution might be another approach to finding the ground state, in addition to the iterative method described in the main text.
Imaginary time evolution might also be a feasible approach to determine $C[l]$ as an alternative to solving Eq.~\ref{eq:static-c}.

\section{Numerical simulation details}
\label{sec:app-numerical}
All numerical quantum circuit simulation is performed using the \textsc{TensorCircuit}~\cite{zhang2023tensorcircuit} package
and the \textsc{TenCirChem}~\cite{li2023tencirchem} package
without considering noise.
Classical DMRG simulation is performed using the \textsc{Renormalizer} package~\cite{renormalizer}.
We use harmonic oscillator eigenstates for phonon basis states.
Using positional states might affect the performance of traditional encodings because of the truncation,
however, we expect variational encoding to be insensitive to the choice of phonon basis states at the $N\rightarrow \infty$ limit.
We use Gray code for binary encoding as an improvement to the standard approach~\cite{sawaya2020resource}.
For both ground state simulation and dynamics simulation, $C[l]$ is initialized as $C[l]_{mn} = \delta_{mn}$.

For the VQE simulation of the Holstein model, 
the circuit parameters $\vec \theta$ are optimized by the L-BFGS-G method implemented in \textsc{SciPy} package~\cite{2020SciPy-NMeth}.
The parameter gradient is calculated by auto-differentiation.
The initial values for the parameters are set to zero at the first round of the macro-iteration.
In subsequent macro-iterations, the previously optimized parameters are used as the initial value for faster convergence.
Eq.~\ref{eq:static-c} is solved by the DF-SANE method implemented in \textsc{SciPy}~\cite{2020SciPy-NMeth}.
Since this is a non-linear equation, we provide three initial guesses and adopt the one with the lowest energy.
The solved $C[l]$ sometimes does not satisfy the orthonormal condition due to numerical imprecision
and the orthonormal condition is enforced by QR decomposition in each macro-iteration.

For the VQD simulation of the spin-boson model, the variational Hamiltonian ansatz used is more complex than the VQE simulation.
Because $C[l]$ is complex, $\hat B[l] \hat h[l]_x \hat B[l]^\dagger$ spans the whole Hermitian matrix space.
Thus for  $\hat h[l]_x$ the whole Pauli matrix set $\{X, Y, Z, I\}^{\otimes N_l}$ is added to the ansatz.
To obtain the quantities required to calculate $\theta_k$, 
the Jacobian of the wavefunction $\phi(\vec \theta)$ is firstly calculated by auto-differentiation,
and then the r.h.s and l.h.s of Eq. 5 in the main text are calculated by matrix multiplication.
How to measure the quantities in realistic quantum circuits is well described in the literature~\cite{lee2022variational}.
To calculate $\dot C[l]$ it is necessary to take the inverse of $\rho[l]$ which is sometimes ill-conditioned.
We add $1\times 10^{-5}$ to the diagonal elements of $\rho[l]$ for regularization.
The time evolution of $\theta_k$ and $C[l]$ is carried out using the RK45 method implemented in \textsc{SciPy}~\cite{2020SciPy-NMeth}.
We observe that the gradient of $\theta_k$ is usually much larger than $C[l]$. Thus it is possible to evolve the two sets of parameters separately, which deserves further investigation.
For Trotterized time evolution, $N=16$ and a time step of 0.01 are used.

\section{Experiments on a superconducting quantum processor}

\subsection{Device parameters}
The superconducting quantum processor, as shown in Fig.~3(a) in the main text, is composed of nine computational transmon qubits with each pair of neighboring qubits mediated via a tunable coupler, forming a cross-shaped architecture~\cite{yan2018tunable, li2020tunable}. Each computational qubit has an independent readout cavity for state measurement and $XY$/$Z$ control lines for state operation. High-fidelity simultaneous single-shot readout for all qubits are achieved with the help of the multistage amplification with the Josephson parametric amplifier (JPA) functioning as the first stage of the amplification. The fundamental device parameters including qubit parameters and gate parameters are outlined in Table.~\ref{tab:SQqubitparameter} and Table.~\ref{tab:TQqubitparameter}, where the parasitic $ZZ$ interaction between qubits is suppressed by the coupler.

\begin{table*}[ht]
\caption{Single qubit gate parameters. $\omega_r$ is the resonant frequency of the readout cavity for each qubit. $\omega_{j,max}\, (j=1 \sim 9)$ are the maximum resonant frequencies when qubits are biased at the sweet spot. $\omega_{j,idle}\, (j=1 \sim 9)$ are the idle frequencies for implementing the single-qubit operations. $\alpha_j \, (j=1 \sim 9)$ are the qubits' anharmonicities. $T_1$, $T_{2,idle}$ and $T_{2E,idle}$ are the corresponding energy relaxation time, Ramsey dephasing time and echoed dephasing time for the qubits measured at the idle frequency. The readout fidelities are typically characterized by detecting each qubit in $\ket{g}$ ($\ket{e}$) when it is prepared in $\ket{g}$ ($\ket{e}$), labeled by $F_{0,j}$ and $F_{1,j}$. To mitigate the error coming from the readout infidelity, the outcomes are reconstructed with the calibration matrix through the Bayes' rule. Single-qubit errors $e_{sq}$ are measured with randomized benchmarking (RB).}

\begin{threeparttable}
\begin{tabular}{cp{1.6cm}<{\centering}p{1.6cm}<{\centering}p{1.6cm}<{\centering}p{1.6cm}<{\centering}p{1.6cm}<{\centering}p{1.6cm}<{\centering}p{1.6cm}<{\centering}p{1.6cm}<{\centering}p{1.6cm}<{\centering}p{1.6cm}<{\centering}}
\hline
\hline
 &{$Q_0$} &{$Q_1$} &{$Q_2$} &{$Q_3$} &{$Q_4$} &{$Q_5$} &{$Q_6$} &{$Q_7$} &{$Q_8$} \tabularnewline
\hline
$\omega_r$ (GHz) &{$6.874$} &{$6.825$} &{$6.931$} &{$6.901$} &{$6.845$} &{$6.786$} &{$6.991$} &{$6.961$} &{$6.806$} \tabularnewline
$\omega_{j,max}$ (GHz) &{$4.003$} &{$4.215$} &{$4.479$} &{$4.689$} &{$4.470$} &{$4.479$} &{$4.657$} &{$4.512$} &{$4.362$} \tabularnewline
$\omega_{j,idle}$ (GHz) &{$3.988$} &{$4.187$} &{$4.464$} &{$4.668$} &{$4.404$} &{$4.359$} &{$4.641$} &{$4.498$} &{$4.223$} \tabularnewline
$\alpha_j/2\pi$ (MHz) &{$-260$} &{$-258$} &{$-255$} &{$-250$} &{$-254$} &{$-258$} &{$-253$} &{$-257$} &{$-264$} 
\tabularnewline
$T_1$ ($\mu$s) &{$35.3$} &{$31.6$} &{$29.5$} &{$27.7$} &{$33.9$} &{$34.3$} &{$33.3$} &{$22.1$} &{$31.8$} \tabularnewline
$T_{2,idle}$ ($\mu$s) &{$11.0$} &{$10.2$} &{$32.6$} &{$38.2$} &{$9.1$} &{$5.6$} &{$43.1$} &{$24.1$} &{$4.3$} \tabularnewline
$T_{2E,idle}$ ($\mu$s) &{$48.2$} &{$38.4$} &{$47.8$} &{$44.2$} &{$31.6$} &{$21.8$} &{$56.8$} &{$32.9$} &{$18.6$} \tabularnewline
$F_{0,j}$ (\%) &{$96.9$} &{$97.4$} &{$98.6$} &{$98.9$} &{$98.7$} &{$98.4$} &{$96.3$} &{$97.2$} &{$94.1$} \tabularnewline
$F_{1,j}$ (\%) &{$93.7$} &{$94.3$} &{$92.5$} &{$94.3$} &{$94.5$} &{$94.6$} &{$92.7$} &{$92.4$} &{$90.9$} \tabularnewline
$e_{sq}$ (\%) &{$0.07$} &{$0.32$} &{$0.06$} &{$0.07$} &{$0.08$} &{$0.05$} &{$0.06$} &{$0.15$} &{$0.08$} \tabularnewline
\hline
\end{tabular} \vspace{0pt}
\label{tab:SQqubitparameter}
\end{threeparttable}
\end{table*}

\begin{table*}[ht]
\caption{Two qubits gate parameters. $\omega_{c,idle}$ are the idle frequencies for each coupler where the $ZZ$ interaction between neighboring computational qubits are maximally suppressed. $\xi_{ZZ}$ is the residual $ZZ$ interaction between each qubit pairs. Two-qubit gates are implemented with the controlled-Z (CZ) and the corresponding gate errors $e_{tq,CZ}$ are characterized with RB.}
\begin{threeparttable}
\begin{tabular}{cp{1.83cm}<{\centering}p{1.83cm}<{\centering}p{1.83cm}<{\centering}p{1.83cm}<{\centering}p{1.83cm}<{\centering}p{1.83cm}<{\centering}p{1.83cm}<{\centering}p{1.83cm}<{\centering}p{1.83cm}<{\centering}}
\hline
\hline
 &{$Q_0-Q_1$} &{$Q_0-Q_2$} &{$Q_0-Q_3$} &{$Q_0-Q_4$} &{$Q_1-Q_5$} &{$Q_2-Q_6$} &{$Q_3-Q_7$} &{$Q_4-Q_8$}\tabularnewline
\hline
$\omega_{c,idle}$ (GHz) &{$5.020$} &{$5.445$} &{$5.570$} &{$5.335$} &{$5.325$} &{$5.595$} &{$5.695$} &{$5.355$} \tabularnewline
$|\xi_{ZZ}|$ (kHz) &{$18.0$} &{$10.0$} &{$5.0$} &{$8.0$} &{$2.0$} &{$3.0$} &{$5.0$} &{$2.0$} \tabularnewline
$e_{tq,CZ}$ (\%) &{$1.57$} &{$2.22$} &{$1.99$} &{$2.47$} &{$0.91$} &{$1.04$} &{$1.2$} &{$0.96$} \tabularnewline
\hline
\end{tabular} \vspace{0pt}
\label{tab:TQqubitparameter}
\end{threeparttable}
\end{table*}

\subsection{Experimental details}
We use three qubits out of the 9-qubit computer for the 2-site Holstein model
\begin{equation}
    \hat H = -V (a^\dagger_1 a_2 + a^\dagger_2 a_1) + \omega b^\dagger_1 b_1
    + \omega b^\dagger_2 b_2
    + g\omega a^\dagger_1 a_1 (b^\dagger_1 + b_1)   + g\omega a^\dagger_2 a_2 (b^\dagger_2 + b_2) \ .
\end{equation}
The electronic degree of freedom is mapped to the second qubit.
Thus, $a^\dagger_1 a_1$ is mapped to $\frac{1}{2}(1 + Z_1)$
and $a^\dagger_2 a_2$ is mapped to $\frac{1}{2}(1 - Z_1)$.
The phonon modes are mapped to the first and the third qubit. 
With binary encoding and $N_l=1$, the Hamiltonian in the Pauli string form reads
\begin{equation}
    \hat H = -V X_1 + \frac{1}{2}\omega (1 - Z_0)+ \frac{1}{2}\omega (1 - Z_2)
    + \frac{1}{2}g\omega (1+Z_1) X_0 + \frac{1}{2}g\omega (1-Z_1) X_2 \ .
\end{equation}
For variational encoding, we assume $C[l] = C$. That is, the two modes share the same variational encoder.
This is a reasonable assumption for translational invariant systems.
\wtli{
$\hat b^\dagger \hat b = \sum_m m \ket{m} \bra{m}$ is then encoded to
\begin{equation}
\begin{aligned}
    \hat B (\hat b^\dagger \hat b) \hat B^\dagger & = \sum_{nn'} F_{nn'} \ket{n}\bra{n'} \ , \\
    F_{nn'} & := \sum_m m C_{mn} C_{mn'} \ .
\end{aligned}
\end{equation}
It is then possible to express the encoded operator as
\begin{equation}
\hat B (\hat b^\dagger \hat b) \hat B^\dagger = c_{1i} I + c_{1x} X + c_{1z} Z  \ ,
\end{equation}
where
\begin{equation}
\begin{aligned}
c_{1i} & = (F_{00} + F_{11}) / 2 \ ,\\
c_{1x} & = F_{01} = F_{10} \ , \\
c_{1z} & = (F_{00} - F_{11}) / 2 \ .
\end{aligned}
\end{equation}
Similarly,  $\hat b^\dagger + \hat b$ is encoded as
\begin{equation}
    \hat B (\hat b^\dagger + \hat b) \hat B^\dagger = c_{2i} I + c_{2x} X + c_{2z} Z
\end{equation}
and we omit the explicit expression for $c_2$ for brevity.
}
The encoded Hamiltonian is then
\begin{equation}
\begin{aligned}
    \hat H & = -V X_1 + \omega (c_{1i} I_0 + c_{1x} X_0 + c_{1z} Z_0)+ \omega (c_{1i} I_2 + c_{1x} X_2 + c_{1z} Z_2) \\
    & \quad + \frac{1}{2}g\omega (1+Z_1) (c_{2i} I_0 + c_{2x} X_0 + c_{2z} Z_0) + \frac{1}{2}g\omega (1-Z_1) (c_{2i} I_2 + c_{2x} X_2 + c_{2z} Z_2) \ .
\end{aligned}
\end{equation}
We use the following ansatz for the parameterized quantum circuit
\begin{equation}
\label{eq:hw-ansatz}
    \ket{\phi} = \prod_{j=1}^2e^{\theta_{j}a^\dagger_j a_j (b^\dagger_j - b_j)} \frac{1}{\sqrt{2}} \left (\ket{000} + \ket{100} \right) \ ,
\end{equation}
Because $C[1] = C[2]$, the parameter space can be further simplified by setting $\theta_1 = \theta_2$.
With binary encoding, the ansatz transforms to
\begin{equation}
    \ket{\phi} =  e^{i\theta Y_2} e^{-i\theta Z_1 Y_2} e^{i\theta Y_0} e^{i\theta Z_1 Y_0} H_1 \ket{0} \ .
\end{equation}
The ansatz is compiled into the following quantum circuit with 4 CNOT gates.

\scalebox{1}{
\Qcircuit @C=1.0em @R=0.2em @!R { \\
	 	\nghost{{q}_{0} :  } & \lstick{{q}_{0} :  } & \gate{R_Z(\frac{-\pi}{2})} & \gate{H} & \targ & \gate{R_Z(-\theta)} & \targ & \gate{R_Z(-\theta)} & \qw & \qw  & \qw & \qw &  \gate{H} & \gate{R_Z(\frac{\pi}{2})} & \qw & \qw\\
	 	\nghost{{q}_{1} :  } & \lstick{{q}_{1} :  } & \gate{H} & \qw & \ctrl{-1} & \qw & \ctrl{-1} & \qw & \ctrl{1} & \qw & \ctrl{1} & \qw & \qw & \qw & \qw & \qw\\
	 	\nghost{{q}_{2} :  } & \lstick{{q}_{2} :  } & \gate{R_Z(\frac{-\pi}{2})} & \gate{H} & \qw & \qw & \qw & \qw & \targ & \gate{R_Z(\theta)} & \targ &\gate{R_Z(-\theta)} & \gate{H} & \gate{R_Z(\frac{\pi}{2})} &  \qw & \qw \\
\\ }}

Each energy term is measured by 8192 shots, 
and the uncertainty is obtained by repeating the measurement 5 times and taking the standard deviation.
For the update of $C[l]$, 4096 shots are performed for each term.
Local readout error mitigation is applied for all results presented unless otherwise stated. 

In Fig.~\ref{fig:vqe} we plot the energy landscape $E(\theta)/V$ in VQE with binary encoding.
Both raw data and data with local readout error mitigation (EM) are presented
for the energy expectation from quantum hardware.
The mitigated landscape is in decent agreement with the perfect simulator.
A minimum at around $\theta = 0.6$ is clearly visible.
We note that the perfect simulator is also based on the $N_l=1$ ansatz and $N$ is far smaller than what is physically demanded.
Thus the minimum presented by the perfect simulator can not be recognized as the ground truth.

\begin{figure}[h]
\includegraphics[width=.35\linewidth]{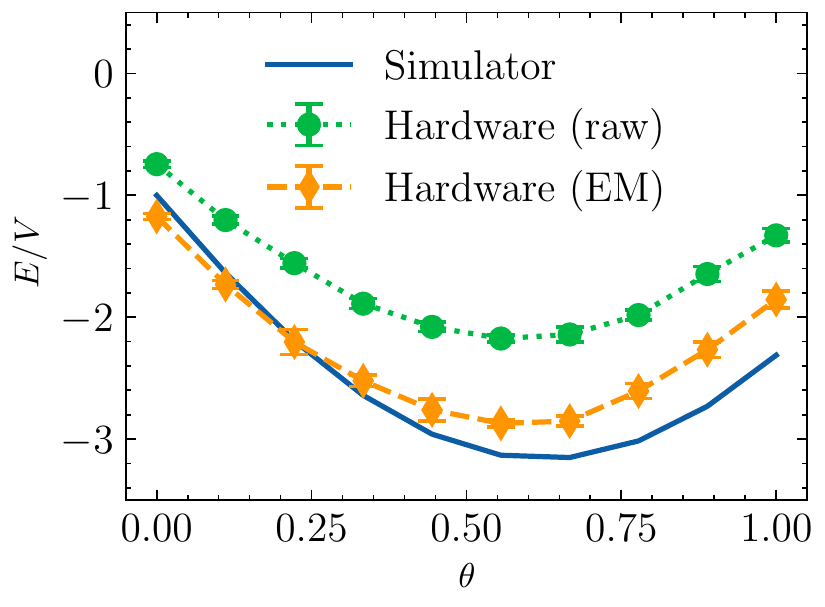}
\caption{\label{fig:vqe} 
VQE energy landscape for the 2-site Holstein model with binary encoding.
For the data from quantum hardware, both raw data and data with readout error mitigation are presented.
The error bar indicates the measurement uncertainty.
}
\end{figure}
\end{widetext}

\bibliography{refs}

\end{document}